\documentclass[onecolumn]{IEEEtran}
\usepackage{pifont}
\usepackage{graphicx}
\usepackage{fancyhdr}
\usepackage{amsmath,amssymb,amstext,amsfonts}
\usepackage{algorithm}
\usepackage{array}
\usepackage{cite}
\usepackage{subfigure}
\usepackage{balance}
\usepackage{algorithmicx}
\usepackage{algpseudocode}

\usepackage{braket}
\usepackage{tabulary}
\usepackage{diagbox}

\begin{document}

\title{Channel Modeling of Satellite-to-Underwater Laser Communication Links: An Analytical-Monte Carlo Hybrid Approach}

\author{Zhixing~Wang, Renzhi~Yuan, \IEEEmembership{Member,~IEEE}, Haifeng~Yao, Chuang~Yang, and Mugen~Peng, \IEEEmembership{Fellow,~IEEE}.
\thanks{Zhixing Wang, Renzhi Yuan, Chuang Yang, and Mugen Peng are with the State Key Laboratory of Networking and Switching Technology, Beijing University of Posts and Telecommunications, Beijing, China (e-mails: wangzhixing@bupt.edu.cn; renzhi.yuan@bupt.edu.cn; chuangyang@bupt.edu.cn; pmg@bupt.edu.cn); Haifeng Yao is with School of Optics and Photonics, Beijing Institute of Technology, Beijing 100081, China, and Yangtze Delta Region Academy of Beijing Institute of Technology, Jiaxing 314019, China (custfeng@outlook.com). A journal version is submitted for peer-review process.}
\thanks{This work is supported by the National Natural Science Foundation of China under Grant No. 62201075.}
\thanks{Corresponding author: Renzhi Yuan}}

\maketitle

\begin{abstract}
Channel modeling for satellite-to-underwater laser communication (StULC) links remains challenging due to long distances and the diversity of the channel constituents. The StULC channel is typically segmented into three isolated channels: the atmospheric channel, the air-water interface channel, and the underwater channel.
Previous studies involving StULC channel modeling either focused on separated channels or neglected the combined effects of particles and turbulence on laser propagation.
In this paper, we established a comprehensive StULC channel model by an analytical-Monte Carlo hybrid approach, taking into account the effects of both particles and turbulence. We first obtained the intensity distribution of the transmitted laser beam after passing through the turbulent atmosphere based on the extended Huygens-Fresnel principle. Then we derived a closed-form probability density function of the photon propagating direction after passing through the air-water interface, which greatly simplified the modeling of StULC links. At last, we employed a Monte Carlo method to model the underwater links and obtained the power distribution at the receiving plane.
Based on the proposed StULC channel model, we analyzed the bit error rate and the outage probability under different environmental conditions. Numerical results demonstrated that, the influence of underwater particle concentration on the communication performance is much pronounced than those of both the atmospheric turbulence and the underwater turbulence.
Notably, increasing the wind speed at the air-water interface does not significantly worsen the communication performance of the StULC links.
\end{abstract}


\section{Introduction}
\subsection{Background and Motivation}
In recent years, with the increasing focus on oceanographic research and underwater commercial development, such as underwater exploration, real-time underwater monitoring video streaming transmission, and the exploitation of underwater natural resources, there has been a growing interest in air-water cross-medium communication. Current underwater communication systems predominantly rely on acoustic or radio frequency systems, connecting with above-water nodes by deploying relays on the sea surface, which are limited by low data transmission rates and high latency \cite{sun2020field}.
Laser communication boasts advantages such as high bandwidth, low latency, and robust security \cite{angara2023underwater}. Moreover, when using laser for air-water cross-medium communication, it enables data transmission within a certain depth range without the need for relay nodes on the water surface \cite{luo2022recent}. Especially, owing to lower absorption of blue-green laser underwater, laser communication can achieve even Gbps data transfer rate at medium ranges \cite{dong2025towards}. These advantages make laser communication a promising option for air-water cross-medium communication. At the same time, satellite communication, with its extensive coverage and immunity to terrain factors \cite{hui2025review}, offers more stable and flexible connections for communication with underwater devices. Therefore, satellite-to-underwater laser communication (StULC) provides extensive and informative connections for fixed as well as mobile devices underwater.

The StULC channel is highly complex and is often divided into three parts: the atmospheric channel, the air-water interface channel, and the underwater channel. In the atmospheric channel, atmospheric attenuation induces energy loss, whereas atmospheric turbulence elicits laser beam wander, spreading, and light intensity scintillations, collectively affecting the transmission characteristics of laser. In the air-water interface channel, refractions by the random rough sea surface exacerbate beam spreading while altering the laser propagation direction, thereby degrading transmission performance. In the underwater channel, turbulence-induced fading and particle-induced absorption and scattering impose an additional layer of energy loss on the laser beam. Given the high complexity of the StULC channel, establishing an comprehensive channel model is of great significance for performance analysis and system design of the StULC systems.

\subsection{Related Works}
\subsubsection{Atmospheric Channel Modeling}
In the atmospheric channel, atmospheric turbulence induces effects such as beam wander, beam spreading and light intensity scintillation, which has long been a research emphasis. Numerous studies have focused on characterizing it using phase screen method, Monte Carlo method and the extended Huygens-Fresnel integral method.
The phase screen method is currently the most widely used numerical simulation technique for modeling the propagation of light waves through the turbulent atmosphere \cite{liu2019improved,paulson2019randomized}. Phase screens are placed layer by layer along the beam propagation path. The statistically generated phase screens at each layer represent the phase that would have occurred in propagation through turbulence from the previous layer. There are many methods for generating phase screen, including the power spectral inversion method \cite{coles1995simulation}, the Zernike polynomial method \cite{noll1976zernike}, and the fractal method \cite{perez2004modeling}, etc. A recent study extended phase screen method from phase domain to amplitude domain \cite{yao2019analysis}. As a numerical method, the phase screen method enables flexible parameter configuration for modeling turbulent media. However, the number and placement of phase screens are determined empirically in most cases.
The Monte Carlo method is also widely employed to model the atmospheric optical channel \cite{song2024study,hu2020link}. The Monte Carlo method performs physical simulation of individual photon trajectories, followed by statistical calculation of the photon reception probability. A link attenuation model of fog channel was established based on Monte Carlo simulation and Kim formula, with its accuracy experimentally validated \cite{hu2020link}. The Monte Carlo method is renowned for its accuracy and flexibility, yet it suffers from low computational efficiency, especially in modeling the long-range satellite-to-ground links.
The extended Huygens-Fresnel integral method \cite{lutomirski1971propagation} was first introduced to address the propagation of plane waves through turbulent media. Then it was combined with the Rytov approximation theory to provide a detailed discussion on the statistical characteristics of laser propagation in turbulent media \cite{andrews2005laser}. The the average intensity of Gaussian beam propagation in the turbulent atmosphere, incorporating the effects of turbulence scale, was derived by an approximation of the extended Huygens-Fresnel integral \cite{wang2007average}. As an analytical method, the extended Huygens-Fresnel integral method allows for quick calculation of statistical properties of light waves in turbulent atmospheres. While the extended Huygens-Fresnel integral method provides mean intensity and variance, it fails to provide the joint PDF of light intensity, which restricts comprehensive channel modeling.

\subsubsection{Air-Water Interface Channel Modeling}
In the air-water interface channel, laser scattering from random rough sea surface has a great influence on application performance. Existing research methods typically involve first modeling the random rough sea surface and then applying Fresnel's Law to compute the refraction of light beams, where the sea surface modeling can be broadly categorized into two classes: the ocean-wave-spectrum-based model and the Cox-Munk model. Ocean wave spectrum can be applied in either the linear superposition method or the linear filtering method to construct three-dimensional sea surface models \cite{su2021ocean}. A three-dimensional wave model driven by the wind was built to obtain the transmittance and refraction angles of the optical path \cite{dong2020fast}. The influences of the roughness of sea surface and wind speed on laser transmission and energy distribution were analyzed \cite{li2018research}. The transmission characteristics of blue-green lasers through two-dimensional dynamic sea surfaces were investigated in the presence of sea surface-bubble layer \cite{Wang2022transmission}. Cox-Munk model \cite{cox1954measurement} is often used to describe the statistical behavior of pitching angle of seawater. The performance of an underwater optical wireless communication system was investigated using both transmitted and reflected beams, and the impact of random motions was obtained experimentally from a water tank \cite{majumdar2012analysis}. The wavy air-water interface was considered in an optical simulation model and the properties of wavy surfaces were explored experimentally \cite{qin2022analysis}. The reflectivity at a given specular point on wavy sea surface was derived based on Cox-Munk model \cite{zhang2021maximum}. The probability density function (PDF) of light spot jitter in underwater-atmosphere uplink optical channel considering refraction of wavy surface was studied by approximating the Cox-Munk model to Rayleigh distribution \cite{ata2024performance}. However, existing research methods based on either the ocean-wave-spectrum-based model or the Cox-Munk model have to rely on Monte Carlo method to obtain laser propagating direction after passing through the air-water interface. If the underwater channel is also based on Monte Carlo method, a huge number of photons is required, which results in extremely low computational efficiency. Therefore, an analytical model for the laser beam refraction at the air-water interface is urgently needed to simplify the channel modeling of air-water interface.

\subsubsection{Underwater Channel Modeling}
In the underwater channel, for studying the impact of underwater particles and turbulence on laser propagation, these methods are frequently utilized: phase screen method and the Monte Carlo method. The underwater phase screen method, similar to its atmospheric counterpart, investigates the influence of turbulence on laser beam from the perspective of optical fields \cite{ji2024generalized}. However, phase screen method is incapable of investigating the scattering effects of seawater and has to be combined with the Monte Carlo method.
The Monte Carlo method is a frequently adopted approach in modeling of underwater channels. An underwater optical channel model in turbulent oceanic clear ocean was established, and the predicted light intensity fluctuations under weak turbulence show good agreement with experimental data \cite{vali2017modeling}. The path loss and channel impulse response of the underwater channel was analyzed and compared by adopting different scattering phase functions \cite{umar2019modelling}. An improved Monte Carlo method based channel model was constructed, which provides a general analysis framework for the absorption and scattering effects brought by the two factors of particles and turbulence \cite{xu2022improvement}.

\subsubsection{Composite Channel Modeling}
While numerous studies have focused on single channels, research on the integrated StULC channel remains limited, with the majority of them employing the Monte Carlo method. The attenuation and time delay spread of laser pulse were analyzed in different channel mediums \cite{dong2009atmosphere}. The spatial distribution characteristics of photons was presented in StULC channel \cite{dong2010monte}. An inhomogeneous atmosphere-to-underwater optical channel was modeled utilizing the Monte Carlo method to investigate the effect of solar radiation noise on the performance of a long-range atmosphere-ocean laser communication system \cite{jiao2025effects}. The impacts of various factors on the reception performance of the satellite-to-underwater optical channel was analyzed using the Monte Carlo method, with a particular emphasis on sea surface wind speed and wave-induced foam \cite{sahoo2022effect}.
Additionally, the transmission characteristics of Gaussian beams in other modes through atmospheric turbulence, air-water interface, and underwater turbulence were studied using the phase screen method \cite{wang2022transmission2}.
However, these researches based on the Monte Carlo method have all overlooked the effects of atmospheric and underwater turbulence on laser beams, while research employing the phase screen method has neglected the influence of underwater particle scattering. Therefore, it is of urgent need to establish a comprehensive channel model for the StULC links taking into account the effects of both particles and turbulence.

\subsection{Contributions}
In this paper, we establish a comprehensive StULC channel model by combining analytical and Monte Carlo methods. First, based on the extended Huygens-Fresnel integral method, we obtain the light intensity distribution of laser beams after propagating through the atmospheric channel. Then by using the Cox-Munk model, we derive the PDF of the zenith angle of the laser propagating direction induced by the random sea surface. Then we employ the Monte Carlo method to model the underwater channel and obtain the power distribution at the underwater receiving plane. At last, based on the proposed StULC channel model, we simulate the communication performance including both the bit-error rate (BER) and the outage probability under various environmental conditions.
The main contributions of this work can be summarized as follows:
\begin{itemize}
\item We established a comprehensive StULC channel model taking into account the effects of atmospheric turbulence, random sea surface, underwater particles-induced scattering and absorption, and underwater turbulence.
\item We proposed an analytical-Monte Carlo hybrid approach for modeling the StULC link by combining the computational advantage of extended Huygens-Fresnel integral method in modeling the long-range atmospheric channel and the accurate advantage of Monte Carlo method in modeling complex underwater channel.
\item We derived a closed-form PDF of the zenith angle of laser propagating direction induced by refractions of the random sea surface based on the Cox-Munk model, which greatly simplified the modeling StULC link.
\item We analyzed the spatial distribution of the underwater receiving power, BER, and outage probability under different environmental conditions. Numerical results revealed that the degradation of communication performance caused by underwater particle concentration substantially exceeds that induced by either atmospheric or underwater turbulence.
\item We also demonstrated that the wind speed at the air-water interface has limited influence on the StULC link performance.
\end{itemize}

The rest of this paper are organized as follows. We first introduce the StULC channel by separately analyzing the atmospheric channel, the air-water interface channel, and the underwater channel in Section~\ref{sysmodel}. Then Section~\ref{ber outage} provides the BER and outage probability of the StULC links. In Section~\ref{results}, numerical results are presented to analyze the communication performance of StULC links. At last, we conclude our work in Section~\ref{conclusion}.

\section{System Model}\label{sysmodel}
\begin{figure}
\centering
\includegraphics[width=0.3\textwidth]{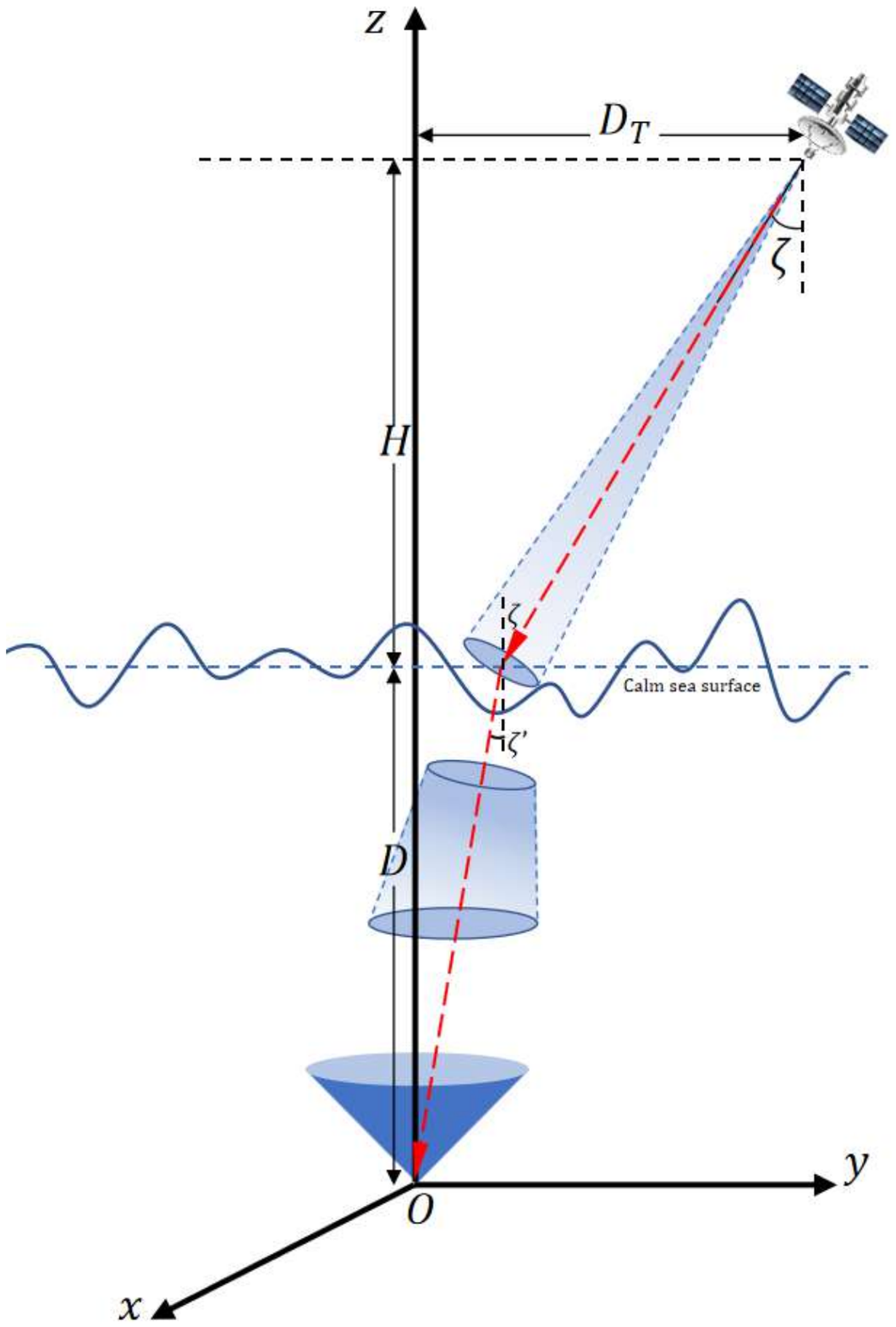}
\caption{System geometry of the StULC channel}
\label{system model figure}
\end{figure}
The system geometry of StULC channel is shown in Fig.~\ref{system model figure}. The laser beam is emitted from the satellite, which is $H$ meters above the sea surface at a certain zenith angle $\zeta$. The receiver is placed at an underwater depth of $D$ meters, located at the origin $(0,0,0)$. Owing to the continuous motion of the satellite, the laser emission angle must be adjusted in accordance with the position of the satellite to align the laser with the target receiver. Since the refractive effect of seawater prevents the laser from directly targeting the underwater object, the zenith angle $\zeta$ at the satellite should be obtained by solving the following equations:
\begin{equation}
\begin{cases}
\eta\sin\zeta=\sin\zeta'\\
D_T=D\tan\zeta'+H\tan\zeta,
\end{cases}
\end{equation}
\noindent where $\zeta'$ is the refraction angle for the calm sea surface, and $\eta=0.75$ is the ratio of the atmospheric refractive index to the seawater refractive index. It is challenging to derive the analytical expression of $\zeta$ directly from the above equations. Therefore, we employ a numerical inversion method to calculate $\zeta$.

\subsection{Atmospheric Channel}
We consider a Gaussian beam in this paper. The Gaussian wave field after propagating a distance $L$ through a turbulent medium can be obtained by the extended Huygens-Fresnel principle, i.e., \cite{andrews2005laser}
\begin{equation}
\begin{aligned}
U(r,L)&=\frac{-\mathrm{i}k\exp(\mathrm{i}kL)}{2\pi L}\int_0^\infty\exp\left[\frac{\mathrm{i}k(r-s)^2}{2L}\right.\\
&\quad\quad\quad\quad\quad\quad\quad\quad\quad\quad +\psi(r,s,L)\Big]U(s,0)\mathrm{d}^2s,
\end{aligned}
\end{equation}
\noindent where $r$ is the beam radius; $k=2\pi/\lambda$ is the optical wave number and $\lambda$ is the wavelength; $\psi(r,s,L)$ is the is the random part of the complex phase of a spherical wave propagating in the turbulent medium; $U(s,0)=U_{0}\exp{\left[-\left(\frac{s^{2}}{2W_{0}^{2}}+\frac{iks^{2}}{2F_{0}}\right)\right]}$ is the wave field at $L=0$, where $W_0=\lambda/(\pi\beta_{T})$ is the effective beam radius and $\beta_T$ is the divergence angle of the laser, $F_0$ is the phase front radius of curvature, $U_0=\sqrt{P_{\mathrm{Tx}}/(\pi W_0^2)}$ is the constant amplitude of the beam, and $P_{\mathrm{Tx}}$ is laser emission power. Then the mean irradiance $\langle I(r,L)\rangle\triangleq \langle |U(r,L)|^2\rangle$ on the wave front at a distance $L$ can be obtained as \cite{andrews2005laser}
\begin{equation}
\begin{aligned}
\langle I(r,L)\rangle =&I_0\left(\frac{k}{2\pi L}\right)^2\iint\!\exp\!\left\{-\mathrm{i}k\rho\left(\frac{s}{F_0}-\frac{s-r}{L}\right)\right\}\\
&\exp\!\left[-\frac{1}{2}D_\psi\left(\rho,L\right)\right]\exp\!\left[-\frac{s^2+\frac{1}{4}\rho^2}{W_0^2}\right]\mathrm{d}^2\rho\mathrm{d}^2s,
\end{aligned}
\end{equation}
\noindent where $I_0=U_{0}^*U_{0}$ and $D_{\psi}\left(\rho,L\right)$ is the wave structure function related to the turbulent atmosphere. Through appropriate approximations, the mean irradiance can be reduced to the Gaussian function as \cite{wang2007average}
\begin{equation}\label{mean_irradiance}
\langle I(r,L)\rangle =\frac{I_{0}W_{0}^{2}}{W_{\mathrm{Lt}}^{2}}\exp\!\left(-\frac{r^2}{W_{\mathrm{Lt}}^2}\right),
\end{equation}
\noindent where $W_{\mathrm{Lt}}$ is the long-term spot radius, which is defined as:
\begin{equation}
W_{\mathrm{Lt}}^2\triangleq W_0^2\left[(\frac{L}{F_0}-1)^2+\frac{L^2}{k^2W_0^4}+\frac{4L^2}{k^2W_0^2\tilde{\rho}_0^2}\right]
\end{equation}
$\tilde{\rho_0}^2=\left[1.46k^2L\int_0^1\left(1-\xi\right)^{5/3}C_n^2\left(\xi H\right)\mathrm{d}\xi\right]^{-6/5}\left[1-0.715\kappa_0^{1/3}\right]^{-1}$ is the coherence length of a spherical wave propagating in the turbulent medium, where $\kappa_0=2\pi/L_0$, $L_0$ is the outer scale of the atmospheric turbulence, and $C_n^2(h)$ is the structure parameter associated with the Hufnagel-Valley model \cite{wang2007average} as a function of altitude $h$ described by
\begin{equation}
\begin{aligned}
C_{n}^{2}(h)=0.00594(w/27)^2(h\times10^{-5})^{10}\exp{(-h/1000)}\\
+2.7\times10^{-16}\exp{(-h/1500)}+C_n^2(0)\exp{(-h/100)},
\end{aligned}
\end{equation}
\noindent where $w=21$ $\mathrm{m/s}$ is the high-altitude wind speed and $C_n^2(0)$ is the nominal value of $C^2_n$ near the sea surface. Without loss of generality, we set $C^2_n(0)=1.7\times10^{-17}\mathrm{m}^{-2/3}$ for weak turbulence, and $C^2_n(0)=1.7\times10^{-13}\mathrm{m}^{-2/3}$ for strong turbulence.

The irradiance after passing through the atmospheric channel can be expressed as
\begin{equation}\label{fluctuation I}
I(r,L)=\langle I(r,L)\rangle\xi_t\xi_f,
\end{equation}
\noindent where $\xi_t$ is the atmospheric transmittance and $\xi_f$ is the turbulent fading coefficient, which can be described by a lognormal distribution with PDF given by
\begin{equation}
f_{\xi_f}(\xi_f)=\frac{1}{\xi_f\sqrt{2\pi\sigma^2_{\ln\xi_f}}}\exp\left(-\frac{\left(\frac{\ln\xi_f}{\mathrm{E}[\xi_f]}+\frac{1}{2}\sigma^2_{\ln\xi_f}\right)^2}{2\sigma^2_{\ln\xi_f}}\right),
\end{equation}
\noindent where $\xi_f>0$, $\mathrm{E}[\xi_f]=1$, and $\sigma^2_{\ln\xi_f}$ is the variance of $\ln\xi_f$.

Without loss of generality, we consider a square receiving plane positioned above the sea surface. The beam axis is normal to the plane and passes through its center, with the side length of the receiving plane sized equal to $2W_{\mathrm{Lt}}$.
The receiving plane is partitioned into a $m\times m$ grid of square cells and each cell is with side length of $d_m$. The grid cells are numbered sequentially from $1$ to $M$ ($M=m^2$), following a standard pattern: starting at the top-left corner (assigned number $1$), proceeding left to right and top to bottom, and ending at the bottom-right corner (assigned number $M$). Besides, we assume that the received irradiance is uniform within each individual grid cell.

It is worth noting that the irradiance fluctuations at each individual grid cells are not statistically independent. Photons close to each other during the propagation are highly likely to experience similar atmospheric turbulent conditions. Therefore, accounting for the turbulent correlation of intensity fluctuations between grid cells is imperative.
Following the grid numbering pattern, we use $\xi_{f,i}$ ($i\in [1,M]$) to describe the turbulent fading coefficient of each cell. Since each $\xi_{f,i}$ conforms to a LN distribution, the joint PDF of $\boldsymbol{\xi_f}\triangleq \left[\xi_{f,1},\xi_{f,2},\cdots ,\xi_{f,M}\right]^{\mathrm{T}}$ accounting for turbulent correlation can be expressed as:
\begin{equation}
f_{\boldsymbol{\xi_f}}\left(\boldsymbol{\xi_f}\right)=\frac{1}{\prod^{M}_{i=1}\xi_{f,i}}\frac{1}{((2\pi)^{M}|\Sigma|)^{\frac{1}{2}}}\cdot\exp\left(-\frac{1}{2}\boldsymbol{z}^\mathrm{T}{\Sigma}^{-1}\boldsymbol{z}\right),
\end{equation}
\noindent where $\boldsymbol{z}=\left\{\ln\xi_{f,1}+\frac{1}{2}\sigma^2_{\ln\xi_{f,1}},\cdots,\ln\xi_{f,M}+\frac{1}{2}\sigma^2_{\ln\xi_{f,M}}\right\}$ and $\mathbf{\Sigma}$ is the covariance matrix of $\ln\boldsymbol{\xi_f}$, which can be expressed as
\begin{equation}
\begin{aligned}
\Sigma=\begin{bmatrix}Cov(\xi_{f,1},\xi_{f,1})&\cdots&Cov(\xi_{f,1},\xi_{f,M})\\
\vdots&\ddots&\vdots\\
Cov(\xi_{f,M},\xi_{f,1})&\cdots&Cov(\xi_{f,M},\xi_{f,M})\end{bmatrix}.
\end{aligned}
\end{equation}

For a downlink path from a satellite, the covariance $Cov(\xi_{f,a},\xi_{f,b})$ between two points $\xi_{f,a}$ and $\xi_{f,b}$ with a distance of $\rho$, can be approximated by \cite{andrews2005laser}
\begin{equation}
\begin{aligned}B(\rho)&\approx{\exp[B_{\ln X}(\rho)+B_{\ln Y}(\rho)]-1}\\
&=\exp\!\left(\frac{\mu_{4d}(\rho)}{\mu_{4d}(0)}\sigma_{\ln X_a}^2\right)\exp\!\left(0.99\left(\frac{k\rho^2\eta_Y}{L}\right)^{\frac{5}{12}}\right.\\
&\quad \quad \quad\quad\quad\times\left.K_{\frac{5}{6}}\left(\sqrt{\frac{k\rho^2\eta_Y}{L}}\right)\sigma_{\ln Y_a}^2\right)-1,
\end{aligned}
\end{equation}
\noindent where $K_v(x)$ is a modified Bessel function of the second kind and $\mu_{4d}(\rho)$ can be expressed as:
\begin{equation}
\mu_{4d}(\rho)=\int_{0}^{1}\frac{C_n^2(\xi H)}{\xi^{\frac{1}{3}}\left(1-\frac{5}{8}\xi\right)^{\frac{7}{5}}}
F_1^1\left(\frac{7}{5};1;\frac{-k\rho^2\eta_X}{8L\xi^{\frac{5}{3}}\left(1-\frac{5}{8}\xi\right)}\right)\mathrm{d}\xi,
\end{equation}
\noindent where $F_1^1(a,c,z)\triangleq \sum_{n=0}^\infty\frac{(a)_n}{(c)_n}\frac{z^n}{n!}$ with $|z|<\infty$ is the confluent hypergeometric function \cite{andrews2005laser}; $\eta_{X}$, $\eta_{Y}$, $\sigma_{\ln X_a}^{2}$, and $\sigma_{\ln Y_a}^2$ are given by
\begin{equation}
\begin{aligned}
&\eta_{X}=\frac{0.92}{1+1.11\sigma_{R,a}^{12/5}},\quad \eta_Y=3(1+0.69\sigma_{R,a}^{12/5}),\\
&\sigma_{\ln X_a}^{2}=\frac{0.49\sigma_{R,a}^2}{\left(1+1.11\sigma_{R,a}^{12/5}\right)^{7/6}},
\sigma_{\ln Y_a}^2=\frac{0.51\sigma_{R,a}^2}{\left(1+0.69\sigma_{R,a}^{12/5}\right)^{5/6}},
\end{aligned}	
\end{equation}
\noindent where $\sigma_{R,a}^2=2.25k^{7/6}\sec^{11/6}(\zeta)\int_{0}^{H}C_{n}^2(h)h^{5/6}\mathrm{d}h$ is the Rytov variance for slant-path propagation.

\subsection{Air-Water Interface Channel Model}
\begin{figure}
\centering
\includegraphics[width=0.2\textwidth]{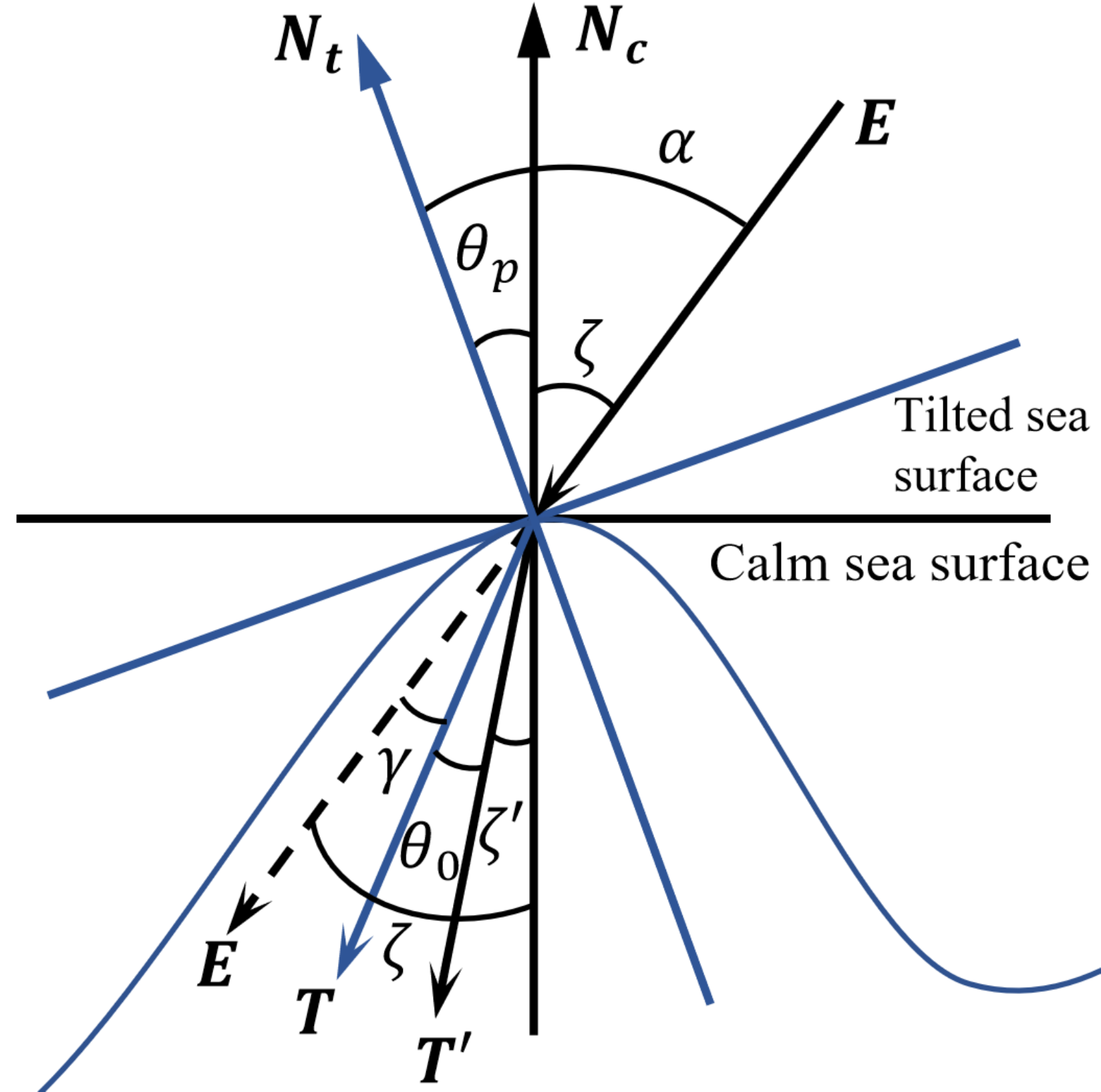}
\caption{Refraction of photons by the rough sea surface}
\label{refraction}
\end{figure}
When a laser beam crosses the air-water interface, it undergoes refraction by seawater. For a calm sea surface, the refraction direction $\boldsymbol{T'}$ is fixed for a given incident direction $\boldsymbol{E}$. However, the wavy sea surface induces stochastic deflections, causing the refracted beam direction to vary randomly around $\boldsymbol{T'}$. 
As shown in Fig.~\ref{refraction}, the incidence angle of the laser beam on the calm sea surface is equal to the emission angle of the laser $\zeta$ and the refraction angle is denoted by $\zeta'$. When the sea surface is tilted, the photon's incidence angle $\alpha=\lvert\zeta+\theta_p\rvert$, where $\theta_p$ is the pitching angle of the sea surface. The Cox-Munk model is widely used to describe the PDF of pitching angle of the sea surface, which is given by  \cite{cox1954measurement}
\begin{equation}\label{cox-munk model}
f_{\theta_p}(\theta_p)=\frac{2}{\sigma^2}\exp\left(\frac{-\tan^2\theta_p}{\sigma^2}\right)\frac{\tan\theta_p}{\cos^2\theta_p}.
\end{equation}
\noindent where $\sigma^2=\sqrt{0.003+0.00512v}$, and $v$ is wind speed above the sea surface. However, the Cox-Munk model does not account for the direction of inclination of the sea surface. Using the calm sea normal vector $\boldsymbol{N}_c$ as the boundary, $\theta_p$ is negative when the tilted sea surface normal vector $\boldsymbol{N}_t$ and the incident beam lie on the same side, and positive when they lie on opposite sides. So we can rewrite \eqref{cox-munk model} as
\begin{equation}
f_{\theta_p}(\theta_p)=\frac{1}{\sigma^2}\exp\left(\frac{-\tan^2\theta_p}{\sigma^2}\right)\frac{\lvert\tan\theta_p\rvert}{\cos^2\theta_p}.
\end{equation}
The PDF of $\alpha$ can be expressed as
\begin{equation}\label{alpha_pdf}
\begin{cases}
f_{\alpha_1}(\alpha)=\frac{1}{\sigma^2}\exp\left(-\frac{\tan^2(\alpha-\zeta)}{\sigma^2}\right)\frac{|\tan(\alpha-\zeta)|}{\cos^2(\alpha-\zeta)},\zeta+\theta_p>0,\\
f_{\alpha_2}(\alpha)=\frac{1}{\sigma^2}\exp\left(-\frac{\tan^2(\alpha+\zeta)}{\sigma^2}\right)\frac{|\tan(\alpha+\zeta)|}{\cos^2(\alpha+\zeta)},\zeta+\theta_p\leq0.
\end{cases}
\end{equation}

Given the unit incident direction vector $\boldsymbol{E}$ and the normal vector of the refractive interface $\boldsymbol{N}_t$, the refracted direction vector $\boldsymbol{T}$ is calculated by
\begin{equation}\label{refracted vector}
\boldsymbol{T}={\eta}\boldsymbol{E}+\boldsymbol{N}_t*\left(\eta\cos\alpha-\sqrt{1-\eta^2(1-\cos^2\alpha)}\right),
\end{equation}
\noindent where $\cos\alpha=-\boldsymbol{N}_t\cdot\boldsymbol{E}$. Obviously, $\gamma$ is the angle between $\boldsymbol{E}$ and $\boldsymbol{T}$, so $\cos\gamma$ can be expressed as a function of $\alpha$, i.e.,
\begin{equation}
\begin{aligned}
\cos\gamma&=\mathbf{E}\cdot \mathbf{T}=\eta\sin^2\alpha+\cos\alpha\times\sqrt{1-\eta^2\sin^2\alpha}.
\end{aligned}
\end{equation}
Then $\alpha$ can be obtained as
\begin{equation}\label{inv_cosgamma}
\alpha=\arcsin\sqrt{\frac{4\left(\cos^2\gamma-1\right)}{3\left(2\cos\gamma-25/12\right)}}.
\end{equation}
As shown in Fig.~\ref{refraction}, $\theta_0$ is the angle between $\boldsymbol{T}$ and $\boldsymbol{T'}$, which can be expressed as
\begin{equation}
\begin{cases}
\theta_0=\lvert\zeta-\zeta'-\gamma\rvert,\quad\,\,\, \zeta+\theta_p>0,\\
\theta_0=\zeta-\zeta'+\gamma,\quad\quad \zeta+\theta_p<0.
\end{cases}
\end{equation}
By substituting $\theta_0=\lvert\zeta-\zeta'-\gamma\rvert$ into \eqref{inv_cosgamma}, we can obtain
\begin{equation}\label{alpha_case1}
\alpha=
\begin{cases}
\arcsin\sqrt{\frac{4\left(\cos^2\left(\zeta-\zeta'+\theta_0\right)-1\right)}{3\left(2\cos\left(\zeta-\zeta'+\theta_0\right)-25/12\right)}},\quad\gamma\geq\zeta-\zeta',\\
\arcsin\sqrt{\frac{4\left(\cos^2\left(\zeta-\zeta'-\theta_0\right)-1\right)}{3\left(2\cos\left(\zeta-\zeta'-\theta_0\right)-25/12\right)}},\,\quad\gamma<\zeta-\zeta'.
\end{cases}
\end{equation}
By substituting $\theta_0=\zeta-\zeta'+\gamma$ into \eqref{inv_cosgamma}, we can obtain
\begin{equation}\label{alpha_case2}
\alpha=\arcsin\sqrt{\frac{4\left(\cos^2\left(\zeta-\zeta'-\theta_0\right)-1\right)}{3\left(2\cos\left(\zeta-\zeta'-\theta_0\right)-25/12\right)}}.
\end{equation}
The PDF of $\theta_0$ can be derived by substituting \eqref{alpha_case1} and \eqref{alpha_case2} into \eqref{alpha_pdf}, which is given in \eqref{pdf_scatter0} at the top of next page, where $J_1(x)=\cos(\zeta-\zeta'+x)$, $J_2(x)=\cos(\zeta-\zeta'-x)$, $h(x)=\arcsin\left(\sqrt{\frac{4(x^2-1)}{6x-25/4}}\right)$, and $\gamma(x)=\arccos\left( \eta\sin^2x+\cos x\times\sqrt{1-\eta^2\sin^2x}\right)$.

\begin{figure*}
\begin{equation}\label{pdf_scatter0}
f_{\theta_{0}}(\theta_{0})=
\begin{cases}
-f_{\alpha_1}\left(h\left(J_{1}(\theta_0)\right)\right)*\frac{\mathrm{d}h}{\mathrm{d}\theta_{0}}+f_{\alpha_1}\left(h\left(J_{2}(\theta_0)\right)\right)*\frac{\mathrm{d}h}{\mathrm{d}\theta_{0}},\quad0\leq\theta_{0}<\zeta-\zeta',\\
-f_{\alpha_1}\left(h\left(J_{1}(\theta_0)\right)\right)*\frac{\mathrm{d}h}{\mathrm{d}\theta_{0}}-f_{\alpha_2}\left(h\left(J_{2}(\theta_0)\right)\right)*\frac{\mathrm{d}h}{\mathrm{d}\theta_{0}},\quad\zeta-\zeta'\leq\theta_{0}<\gamma(\frac{\pi}{2}-\zeta),\\
-f_{\alpha_1}\left(h\left(J_{1}(\theta_0)\right)\right)*\frac{\mathrm{d}h}{\mathrm{d}\theta_{0}},\quad\gamma(\frac{\pi}{2}-\zeta)\leq\theta_0\leq\left|\zeta-\zeta'-\arccos\eta\right|.
\end{cases}
\end{equation}
\end{figure*}

\subsection{Underwater Channel Model}
\subsubsection{Photon Emission}
We employ the Monte Carlo method to model the underwater channel. In atmospheric channel modeling, the laser beam propagates through the turbulent atmosphere until arrivals at a designated receiving plane. Calculating a $m\times m$ mean irradiance matrix $I_{avg}$ on the receiving plane by \eqref{mean_irradiance}, we implement a discrete space sampling approach to generate stochastic photon positions.
The mean irradiance matrix $I_{avg}$ is first normalized to construct a corresponding probability distribution matrix $P_{avg}$, which can be obtained as
\begin{equation}
P_{avg}=\frac{I_{avg}}{\sum_{i=1}^m\sum_{j=1}^mI_{avg}(i,j)}.
\end{equation}
Then $P_{avg}$ is flattened into a column vector $P'_{avg}$ in column-major order, followed by computation of the cumulative sum for each element to construct the cumulative distribution function (CDF), which can be expressed as
\begin{equation}
\mathrm{cdf}(k)=\sum_{i=1}^kP'_{avg}(i),\quad k=1,2,\ldots,m\times m
\end{equation}
To generate $M_p$ photons, we first produce $M_p$ uniformly distributed random numbers $\xi_p\in [0,1]$. For all photons, the random number is then mapped to its corresponding interval index $k_p$ through the CDF. The one-dimensional index $k_p$ is subsequently converted to two-dimensional matrix index $(k_{row,p},k_{col,p})$ using the relation:
\begin{equation}
\begin{cases}
k_{\mathrm{row},p}=((k_p-1)\,\mathrm{mod}\,m)+1,\\
k_{\mathrm{col},p}=\left\lfloor\frac{k_p-1}{m}\right\rfloor+1.
\end{cases}
\end{equation}
With the center of the receiving plane as the origin, after determining the grid cell $(k_{row,p},k_{col,p})$ to which each photon belongs, we introduce random numbers to randomize the photon's position within the cell. Thus, in the coordinate system of the receiving plane, the photon's position $(x_r,y_r)$ is calculated by
\begin{equation}
\begin{cases}
x_{r}=\frac{2W_{\mathrm{Lt}}}{m}\left(k_{\mathrm{col},p}-1+\xi_{x,r}\right)-W_{\mathrm{Lt}},\\
y_{r}=\frac{2W_{\mathrm{Lt}}}{m}\left(k_{\mathrm{row},p}-1+\xi_{y,r}\right)-W_{\mathrm{Lt}},
\end{cases}
\end{equation}
\noindent where $\xi_{x,r}$ and $\xi_{y,r}$ are both uniformly distributed random variable in $[0,1]$. Due to the fluctuations of irradiance, we calculate a $m \times m$ instantaneous irradiance matrix $I_{fluc}$ by \eqref{fluctuation I} and then normalize it according to the spatial weighting matrix $P_{fluc}$. By interpolating the weight matrix, we can obtain the weight of photons $p_i=P_{fluc}(x_{r,i},y_{r,i}),i=1,2,\dots,M_p$, where $p_i$ is the weight of the $i$th photon and $(x_{r,i},y_{r,i})$ is the corresponding photon position.
However, while the receiving plane is perpendicular to the laser beam axis, the oblique incidence of laser on the sea surface means the receiving plane is not parallel to sea level. Actually, when the laser propagates through an slant path and reaches the sea surface, the spot is not circular but elliptical. As illustrated in Fig. \ref{system model figure}, the emission direction vector of the laser $\boldsymbol{E}_T=(0,-\sin\zeta,-\cos\zeta)$. Therefore, the spot is stretched in the $y$-axis direction, and coordinates on the $xOy$ plane of contact points of the photons located at $(x_s,y_s)$ on the receiving plane with the sea surface can be approximately expressed as $(x_s,y_s)=(x_r,y_r\sec\zeta)$.

Due to stochastic waves of the rough sea surface, we need to model the three-dimensional sea surface to determine the $z$-coordinate corresponding to each photon. We model the random fluctuations of the sea surface as a stationary random process, and the correlation function of heights between two arbitrary points on a rough sea surface and the wave spectrum satisfy a Fourier transform relationship under this framework. We employ the linear filtering method for sea surface modeling, which fundamentally operates by performing Fourier transform on white noise, applying ocean wave spectrum filtering in the frequency domain, and then executing inverse Fourier transform to generate the sea surface elevation field.
A three-dimensional sea surface is generated with lengths $L_x=L_y=20$ m along the $x$ and $y$ direction, the number of equidistant discrete points $M_x=M_y=1000$, and the distance between adjacent points $\Delta_x=\Delta_y=0.02$ m. Taking the height of calm sea surface as the zero point of $z$-axis, the height at point $(x,y)$ on the sea surface can be calculated by \cite{pierson1964proposed}
\begin{equation}
\begin{aligned}
f\left(x,y\right)=\frac{1}{L_{x}L_{y}}\sum_{m=-M_x/2}^{M_x/2}\sum_{n=-M_y/2}^{M_y/2}\!&F(\kappa_m,\kappa_n)\\
&\,\,\times\exp\!\left(i\kappa_m x+i\kappa_n y\right),
\end{aligned}
\end{equation}
\noindent where $\kappa_m\frac{2M_x\pi}{L_{x}}, \kappa_n=\frac{2M_y\pi}{L_{y}}$. $F(\kappa_m,\kappa_n)$ can be obtained as
\begin{equation}
\begin{aligned}
F(\kappa_m,\kappa_n)=\left|G(\kappa_m,\kappa_n)\right|S(\kappa_m,\kappa_n)&\left(\sin\angle(G(\kappa_m,\kappa_n))\right.\\
&\left.+\mathrm{i}\cos\angle(G(\kappa_m,\kappa_n))\right),
\end{aligned}
\end{equation}
\noindent where $|G(\kappa_m,\kappa_n)|$ is the amplitude spectrum of noise and $\angle(G(\kappa_m,\kappa_n))=Re\{\ln\left({G(\kappa_m,\kappa_n)}/{\left|G(\kappa_m,\kappa_n)\right|}\right)\}$ is the phase spectrum of noise. $G(\kappa_m,\kappa_n)$ can be calculated as
\begin{equation}
\begin{aligned}
G(\kappa_m,\kappa_n)=\sum_{k=0}^{M_x-1}\sum_{l=0}^{M_y-1}\!&\left(G_{real}(k,l)+\mathrm{i}G_{imag}(k,l)\right)\\
&\quad\quad\exp\!\left(-\mathrm{i}2\pi\left(\frac{k\kappa_m}{M_x}+\frac{l\kappa_n}{M_y}\right)\right),
\end{aligned}
\end{equation}
where $G_{real}(k,l)$ and $G_{imag}(k,l)$ are both random numbers that follow a standard normal distribution. Based on the above formulation, we can obtain the rough sea surface elevation offset relative to a calm sea surface.
When the underwater receiver is located at the origin, accounting for the relative position between photons and the receiver, the coordinates of the contact points between photons and the sea surface $(x_p,y_p,z_p)$ can be obtained as $(x_p,y_p,z_p)=(x_{s},y_{s}+D\tan\zeta',f(x_{s},y_{s})+D)$.

To investigate the impact of the rough sea surface on the propagation direction of photons, we subdivide the rough sea surface into numerous facet elements along the $x$ and $y$-axes. When the facet area becomes sufficiently small, each elemental surface can be effectively approximated as a smooth planar surface treated as refractive interfaces for computation of beam refraction.
The unit normal vector of the facet element $\boldsymbol{N}(n_x,n_y,n_z)=(\cos\phi_p\sin\theta_p,\sin\phi_p\sin\theta_p,\cos\theta_p)$, where $\phi_p$ is the azimuth angle uniformly distributed in $[0,2\pi]$ and $\theta_p$ is the zenith angle which can be obtained by sampling \eqref{cox-munk model}: $\theta_p=\arctan(\sqrt{-\sigma^2\xi_s})$, where $\xi_s$ is a uniform distributed random variable in $[0,1]$.
All photons are approximate to have identical incident direction vectors $\boldsymbol{E}_T$, while being refracted by different sea surface elements. Unit direction vectors of refracted photons $\boldsymbol{T}_R$ can be calculated by \eqref{refracted vector}.

According to the Fresnel's law, if the s-polarized and parallel p-polarized components of a Gaussian laser beam incident on the sea surface have equal intensity magnitudes, the transmittance $Tr$ of the sea surface can be expressed as
\begin{equation}\label{transmittance}
Tr=\frac{1}{2}\left|\frac{\sin2\alpha\sin2\beta}{\sin^2\left(\alpha+\beta\right)}\times\frac{1+\cos^2\left(\alpha-\beta\right)}{\cos^2\left(\alpha-\beta\right)}\right|,
\end{equation}	
\noindent where $\alpha$ and $\beta$ are the angle of incidence and the angle of refraction.

\subsubsection{Photon Propagation}
\begin{figure}
\centering
\includegraphics[width=0.3\textwidth]{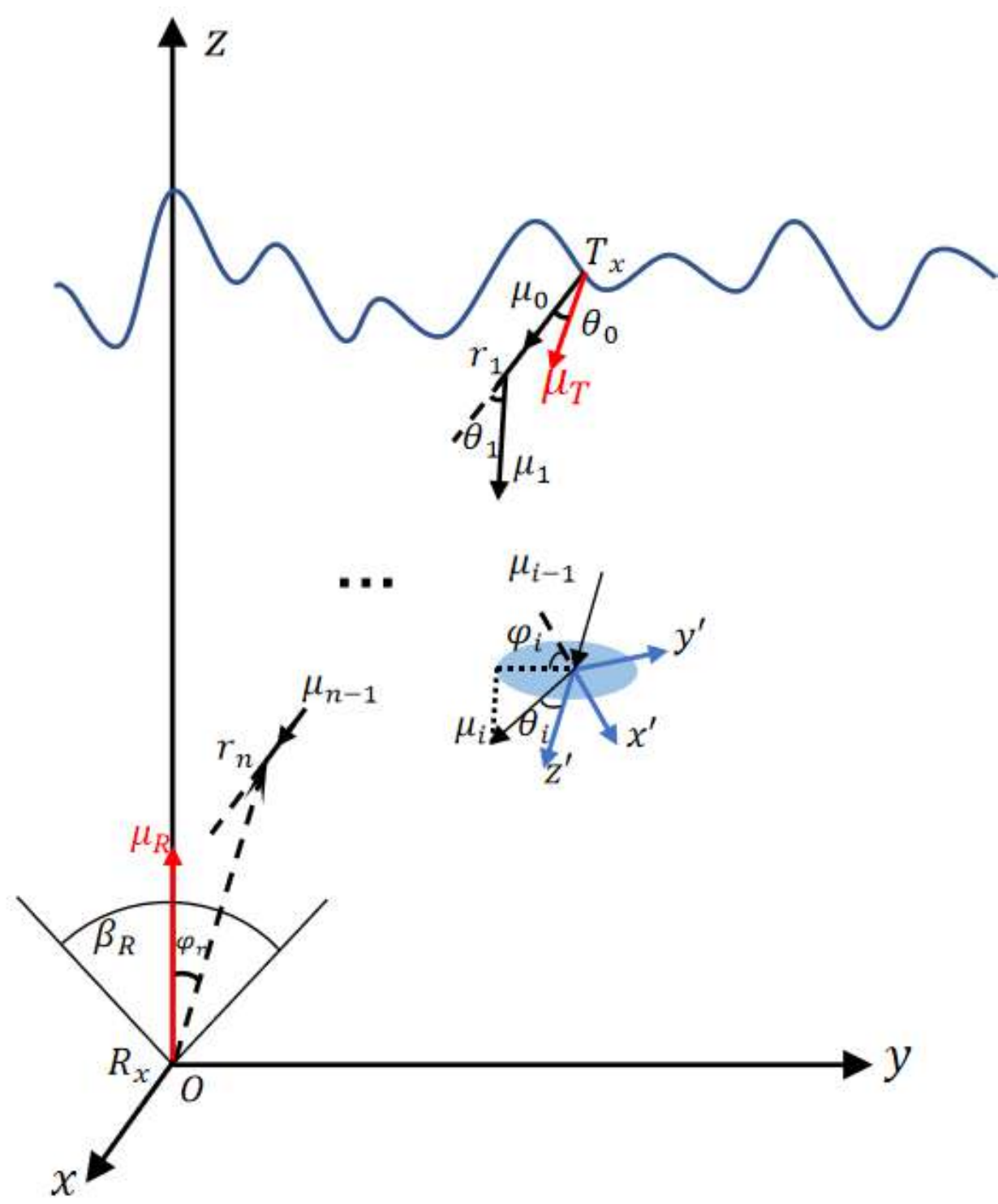}
\caption{Geometric parameters for the underwater propagation process}
\label{Geometric parameter}
\end{figure}
The geometric parameters for the photon propagation process underwater are shown in Fig.~\ref{Geometric parameter}. The receiver Rx is located at the origin $(0,0,0)$. Photons' transmitter positions Tx in the underwater optical channel are their contact points $(x_p,y_p,z_p)$ with sea surface.
The zenith angle, the azimuth angle and the field of view (FOV) angle of the receiver denoted by $\theta_R$, $\phi_R$ and $\beta_R$, respectively. The direction cosines of photons refracted by the calm sea surface and the FOV axis are denoted by $\boldsymbol{\mu_T}$ and $\boldsymbol{\mu_R}$, respectively.

The initial emitting process is regarded as the $zero$-order scattering. We denote the propagating distance, scattering zenith angle and scattering azimuth angle of $i$-order scattering, where $i = 0,1,\dots, n$, by $d_i$, $\theta_i$ and $\phi_i$, respectively. The position of the $i$th scattering point is denoted by $r_i$. The direction cosine of i-order scattering is denoted by $\boldsymbol{\mu_i}$. The scattering zenith angle $\theta_i$ and the scattering azimuth angle $\phi_i$, where $i=1,2,\dots,n$, are the zenith angle and the azimuth angle between $\boldsymbol{\mu_i}$ and $\boldsymbol{\mu_{i-1}}$, while $\theta_0$ and $\phi_0$ is the zenith angle and the azimuth angle of $\boldsymbol{\mu_0}$ and $\boldsymbol{\mu_T}$.

The spatial position of the photon is represented by $(x_i,y_i,z_i)$, and its propagation direction is defined by the directional vector $(\mu_{x,i},\mu_{y,i},\mu_{z,i})$. Photons departing from the air-water interface channel instantaneously enter the underwater optical channel, with the initial position $(x_0,y_0,z_0)=(x_p,y_p,z_p)$ and the initial directional vector $(\mu_{x,0},\mu_{y,0},\mu_{z,0})=\boldsymbol{T}_R$.
For the $zero$-order scattering, when the sea surface fluctuates randomly, the direction of refracted photon will vary around $\mu_T$. The PDF of initial zenith angle is described by \eqref{pdf_scatter0}. The PDF for the initial scattering azimuth angle $\phi_0$ can be obtained as $f(\phi_0)=\frac{1}{2\pi}, 0<\phi_0<2\pi$.
During the propagation, the beam energy loss is primarily caused by the scattering and absorption effects of underwater suspended particles, as well as the fading effects induced by turbulence.
The essence of underwater turbulence is the variation in the refractive index of the medium caused by the non-uniform distribution of water temperature and salinity\cite{baykal2022underwater}, which means the turbulent medium typically lacks significant absorption characteristics in contrast to absorptive media such as suspended particles. Therefore, Similar to atmospheric turbulence, we consider underwater turbulence to only induce intensity fluctuations while neglecting its effects on optical absorption and scattering.
The total attenuation can be described by extinction coefficient $k_e=k_a+k_s$, where $k_a$ is particles-induced absorption coefficient, $k_s$ is particles-induced scattering coefficient.
Between two adjacent scatterings with particles, the photon will travel a random distance $\Delta s$ called step size, which can be expressed as $\Delta s=-{\ln \xi_s}/{k_e}$, where $\xi_s$ is a uniform distributed random variable in [0,1].
For a photon located at $(x_i,y_i,z_i)$ traveling a distance $\Delta s$ in the direction $(\mu_{x,i},\mu_{y,i},\mu_{z,i})$, its spatial position coordinates are updated by:
\begin{equation}
\begin{cases}
x_{i+1}=x_{i}+\mu_{x,i}\Delta s,\\
y_{i+1}=y_{i}+\mu_{y,i}\Delta s,\\
z_{i+1}=z_{i}+\mu_{z,i}\Delta s.
\end{cases}
\end{equation}	

When a photon is scattered by any medium, it continues to propagate in random directions. $f(\phi_i)=\frac{1}{2\pi}$ is the PDF for the scattering azimuth angle $\phi_i$; $f(\theta_i)$ is the PDF for the scattering zenith angles $\theta_i$. If scattered by particles, $f(\theta_i)$ can be approximated as the Henyey-Greenstein (HG) function \cite{gabriel2012monte}
\begin{equation}
f_p(\theta_i)=\frac{1-g^2}{2\pi(1+g^2-2g\cos\theta_i)^{{3}/{2}}},\quad0\leq\theta_i\leq\pi
\end{equation}
\noindent where the constant $g$ is a model parameter obtained from experimental results \cite{gabriel2012monte}.

For the influence of underwater turbulence, we apply the irradiance fluctuation theory to each LOS path between two adjacent particle scatterings, including the path from the last scattering center to the Rx. Then we accumulate the turbulence effects along all the paths to the receiver to describe the probability distribution of received irradiance.
Here, we use the lognormal distribution to describe the normalized irradiance scintillation effect after each LOS path. Under weak oceanic turbulence, the scintillation index, which is so-called Rytov variance for underwater propagation denoted by $\sigma^2_{R,u}$, is defined as \cite{luan2019scintillation}:
\begin{equation}
\sigma^2_{R,u}=8\pi^{2}k^{2}L_u\int_{0}^{1}\!\int_{0}^{\infty}\!\kappa\Phi_{n}(\kappa)\!\left[1-\cos\left(\frac{L_u\kappa^{2}\xi}{k}\right)\right]\mathrm{d}\kappa\mathrm{d}\xi,
\end{equation}
\noindent where $\Phi_n(\kappa)$ is the approximate oceanic refractive-index spectrum with the variable eddy diffusivity ratio \cite{luan2019scintillation}, and $L_u$ is the distance between two adjacent scattering center. Under moderate-to-strong oceanic turbulence, $1\ll \sigma^2_{R,u}$, the scintillation index is defined by \cite{baykal2022underwater}:
\begin{equation}
\sigma^2_{I}=\exp(\sigma^2_{\ln X_u}+\sigma^2_{\ln Y_u})-1,
\end{equation}
\noindent where $\sigma^2_{\ln X_u}=\frac{0.49\sigma_{R,u}^2}{\left(1+1.11\sigma_{R,u}^{12/5}\right)^{7/6}}$ and $\sigma^2_{\ln Y_u}=\frac{0.51\sigma_{R,u}^2}{\left(1+0.69\sigma_{R,u}^{12/5}\right)^{5/6}}$ are called large and small-scale log-irradiance variances, respectively.

When turbulence is considered, the probability of a photon reaching each scattering center along any propagation path is measured by the random irradiance. Each time a photon reaches a scattering center, it travels a LOS path, and the arrival probability undergoes a random fluctuation that follows a LN distribution. Therefore, the probability of a photon reaching the $n$th scattering center follows a distribution that is the multiplicative product of the previous $n$ LN distributions.

Given the direction vector $\boldsymbol{\mu}=(\mu_{x},\mu_{y},\mu_{z})$ of the photon arriving at a scattering center and the deflection angle $(\theta,\phi)$, the relation between new direction vector $\boldsymbol{\mu^{new}}$ and $\boldsymbol{\mu}$ can be derived by \cite{yuan2019monte}
\begin{equation}
\begin{bmatrix}\mu^{new}_x\\\mu^{new}_y\\\mu^{new}_z\end{bmatrix}=\cos\theta\begin{bmatrix}\mu_x\\\mu_y\\\mu_z\end{bmatrix}+\frac{\sin\theta}{\sqrt{1-\mu^2_{z}}}\begin{bmatrix}\mu_x\mu_z\cos\phi-\mu_y\sin\phi\\\mu_{y}\mu_z\cos\phi+\mu_x\sin\phi\\-\cos\phi\end{bmatrix}.
\end{equation}
If $\sqrt{1-\mu_{z}^2}=0$, then
\begin{equation}
\begin{bmatrix}\mu^{new}_x\\\mu^{new}_y\\\mu^{new}_z\end{bmatrix}=\begin{bmatrix}\sin\theta\cos\phi\\\sin\theta\sin\phi\\\mu_z\cos\theta\end{bmatrix}.
\end{equation}

\subsubsection{Photon Detection}
The detection probability after the photon leaving from the $n$th scattering position is given by \cite{yuan2019monte}
\begin{equation}
\begin{aligned}
p_{d}&=I_{n}\frac{k_{s}}{k_{e}}e^{-k_{e}d_{n}}\cos\phi_{r}\int_{\Omega_{r}}f_{\Theta}(\theta_{n})f_{\Phi}(\phi_{n})\mathrm{d}\theta_{n}\mathrm{d}\phi_{n}\\
&\approx I_{n}\frac{k_{s}}{k_{e}}e^{-k_{e}d_{n}}\cos\phi_{r}\min\left(1,f_p(\theta_{n})\Omega_{r}\right)
\end{aligned}
\end{equation}
\noindent where $d_n$ is the distance between $r_n$ and the receiver; $\phi_r$ is the angle between $\boldsymbol{\mu_R}$ and $-\boldsymbol{\mu_n}$; $\Omega_r$ is the solid angle formed by $r_n$ and the receiving area $A_r$; and we have $\Omega_r\approx2\pi\left(1-{d_n}/\sqrt{d_n^2+r_A^2}\right)$, where $r_A=\sqrt{A_r/\pi}$ is the radius of the receiving area; $I_n$ is an indicator function, $I_n=1$ if the $n$th scattering position is located within FOV, $I_n=0$ under other circumstances.
Finally, we obtain the photon arrival probability after $n$ times scattering as
\begin{equation}
P_n=\int\int...\int_Vp_d\left(\frac{k_s}{k_e}\right)^{n-1}\mathrm{d}Q_0\mathrm{d}Q_1...\mathrm{d}Q_{n-1}.
\end{equation}
\noindent where $V$ is the whole integration volume with dimension 3$n$.
For the $zero$-order scattering, the detection probability can be expressed as
\begin{equation}
P_0\approx I_ne^{-k_ed_0}\cos\phi_r\min(1,f_{\theta_0}(\theta_n)\Omega_r).
\end{equation}
\noindent where $d_0$ is the distance between the Tx and the Rx.

The received power after $n$th scattering can be expressed as
\begin{equation}
P_{r,n}=\frac{P_{\mathrm{Tx}}\xi_t}{M}\sum_{m=1}^{M_p}p_m Tr_m P_{m,n},
\end{equation}
\noindent where $Tr_m$ is the transmittance of the sea surface of the $m$th photon and $P_{m,n}$ is the arrival probability of the $m$th photon after $n$ times scattering. The total received power can be obtained as
\begin{equation}\label{P_r}
P_{r}=\sum_{n=0}^{N_s}P_{r,n},
\end{equation}
\noindent where $N_s=4$ is the number of simulated scattering times.

We characterize the turbulence-induced fluctuation effects in the composite channel by turbulent scintillation coefficient $\sigma^2_{tur}$, which can be expressed as \cite{yuan2024monte}
\begin{equation}
\sigma^2_{tur}=\sum_{n=0}^N\left(\frac{P_{r,n}}{P_r}\right)^2\sigma_{tur,n}^2,
\end{equation}
\noindent where $\sigma_{tur,n}^2$ is the turbulent variance for the $n$th scattering order. It can be obtained as
\begin{equation}\label{sturn}
\sigma_{tur,n}^2=\frac{1}{Count}\sum_{m=1}^{M_p}\left[\prod_{i=-1}^nM_2(d_i^m)-1\right]I_n(s_n^m),
\end{equation}
\noindent where $M_p$ is the number of photons, $s^m_n$ denotes the $n$th scattering order center for the $m$th photon, and $Count \triangleq \sum_{m=0}^{M_p}I_n(s_n^m)$ is the number of scattering positions located within FOV. $M_2(d_i^m)=\exp\left(\sigma_R^2(d_i)\right)$ when atmospheric-turbulence-induced and underwater-turbulence-induced fluctuations are both modeled as a LN distribution. $\sigma^2_R$ is the Rytov variance, which is given by $\sigma^2_R=\sigma^2_{R,a}$ when $i=-1$ and $\sigma^2_R=\sigma^2_{R,u}$ at other times. When $i=-1$, $d_{-1}=L$ is the laser propagation distance in the atmospheric optical channel.

\section{BER and Outage Probability}\label{ber outage}
\subsection{BER}A photodetector are employed as the receiver, which transform the light into electrical currents.
Considering the intensity-modulated/direct-detection (IM/DD) with on-off keying (OOK) modulation, the photodetector current of the received signal is
\begin{equation}
i_r=s_kR_dP_r+n_t,
\end{equation}
\noindent where $s_k \in \{0,1\}$ is the $k$-th OOK modulation symbol, $R_d$ is the detector responsivity, $P_r$ is the received power of the receiver obtained in \eqref{P_r}, and $n_t$ a additive white Gaussian noise (AWGN). The received power $P_r$ is insufficiently large so that thermal noise dominates.

By averaging the conditional BER over the received signal's PDF, the mean BER for the OOK system can be expressed as
\begin{equation}
P_e=\int_0^\infty P_e(e|P_r)f_{P_r}(P_r)\mathrm{d}P_r.
\end{equation}
Here, $P_e(e|P_r)$ is the conditional BER, which can be obtained by
\begin{equation}
P_e(e|P_r)=\mathrm{Q}\left(\frac{R_dP_r}{2\sigma_n}\right)=\mathrm{Q}\left(\frac{R_dP_r}{\sqrt{2N_0}}\right),
\end{equation}
\noindent where $\sigma^2_n=N_0/2$ is the variance of $n_t$; $N_0$ is the thermal noise power spectrum defined by $N_0=4KTB/R$, where $K$ is the Boltzmann constant, $R=1\mathrm{M\Omega}$ is the the receiver resistance, $T=300\mathrm{K}$ is temperature in Calvin and $B=1\mathrm{GHz}$ is the noise equivalent bandwidth of the photodetector; $\mathrm{Q}(\cdot)$ is the Gaussian Q function and $f_{P_r}(P_r)$ is the PDF of the received power $P_r$, of which the lognormal distribution demonstrates good fitting performance:
\begin{equation}\label{lognormal pdf}
f_{P_r}(P_r)=\frac{1}{\sigma_R\sqrt{2\pi}}\frac{1}{P_r}\exp\left\{-\frac{\left[\ln\left(\frac{P_r}{\langle P_r\rangle}\right)+\frac{1}{2}\sigma^2_{tur}\right]^2}{2\sigma^2_{tur}}\right\}
\end{equation}

\subsection{Outage Probability}
The SNR $\gamma_s$ associated to each receiver can be obtained as
\begin{equation}\label{SNR}
\gamma_s=\frac{\langle i_r^2\rangle }{\sigma^2_n}=\frac{R_d^2P_r^2}{{N_0}/{2}}=\frac{2R R_d^2P_r^2}{4KTB}.
\end{equation}

Outage probability $P_{\mathrm{out}}$ is an important performance criterion to evaluate communication links. It is defined as the probability that the instantaneous SNR $\gamma_s$ falls below a specified threshold $\gamma_{th}$:
\begin{equation}\label{outage probability define}
P_{\mathrm{out}}=\Pr(\gamma_s<\gamma_{\mathrm{th}}).
\end{equation}
By substituting \eqref{SNR} into \eqref{outage probability define} and setting $\gamma_{th}'=\sqrt{2\gamma_{th}KTB/R_d^2R}$, the outage probability can be expressed as:
\begin{equation}\label{outage probability}
\begin{aligned}
P_{\mathrm{out}}(\gamma_{th})&=\Pr(P_r<\gamma_{th}')=\int_0^{\gamma_{th}'}f_{P_r}(P_r)\mathrm{d}P_r.
\end{aligned}
\end{equation}
By substituting \eqref{lognormal pdf} into \eqref{outage probability}, the above equation can be rewritten as:
\begin{equation}
P_{\mathrm{out}}(\gamma_{th})=1-\mathrm{Q}\left(\frac{\ln\left(\sqrt{\frac{2\gamma_{th}KTB}{\langle P_r\rangle^2R_d^2R}}\right)+\frac{1}{2}\sigma^2_{tur}}{\sigma_{tur}}\right).
\end{equation}
\section{Numerical Results}\label{results}

\begin{figure*}
\begin{center}
\subfigure[]{\includegraphics[width=0.3\textwidth]{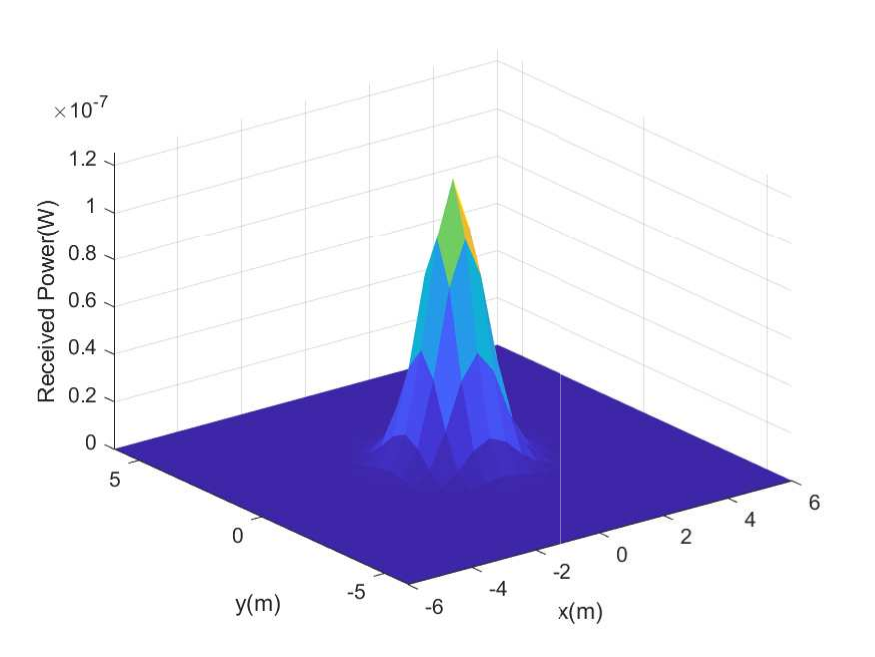}
\label{p_compare}}
\subfigure[]{\includegraphics[width=0.3\textwidth]{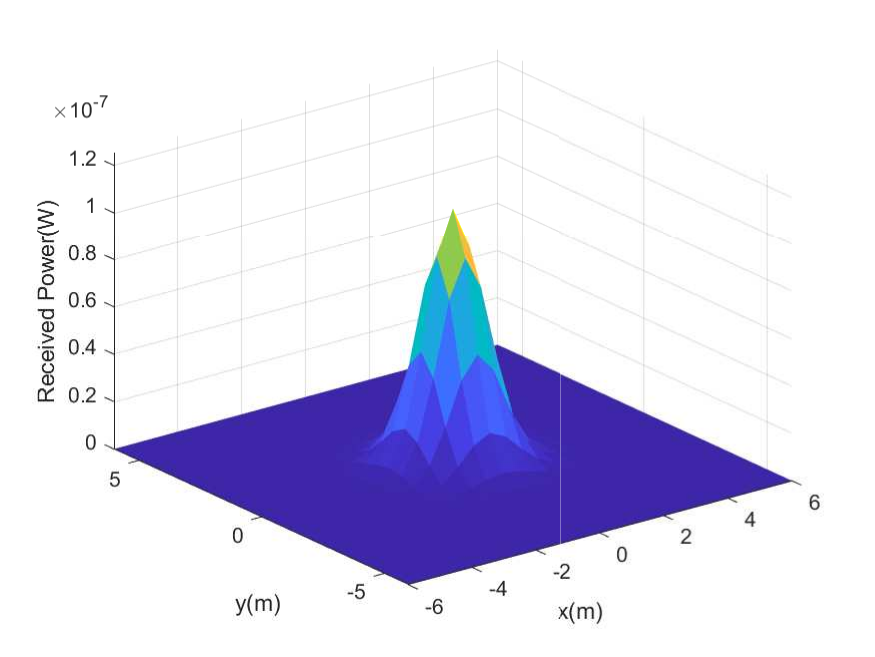}
\label{p_wind}}
\subfigure[]{\includegraphics[width=0.3\textwidth]{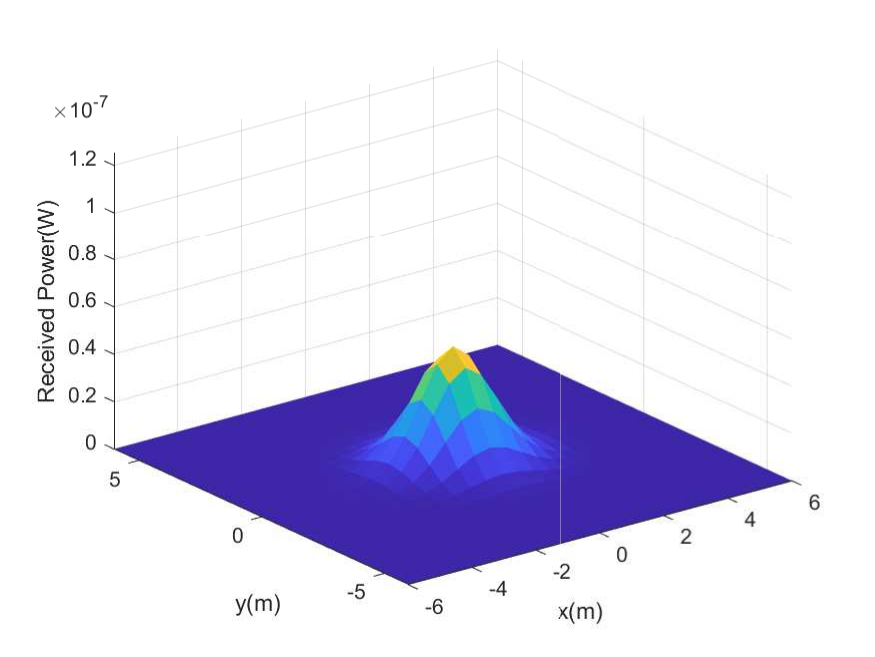}
\label{p_angle}}
\subfigure[]{\includegraphics[width=0.3\textwidth]{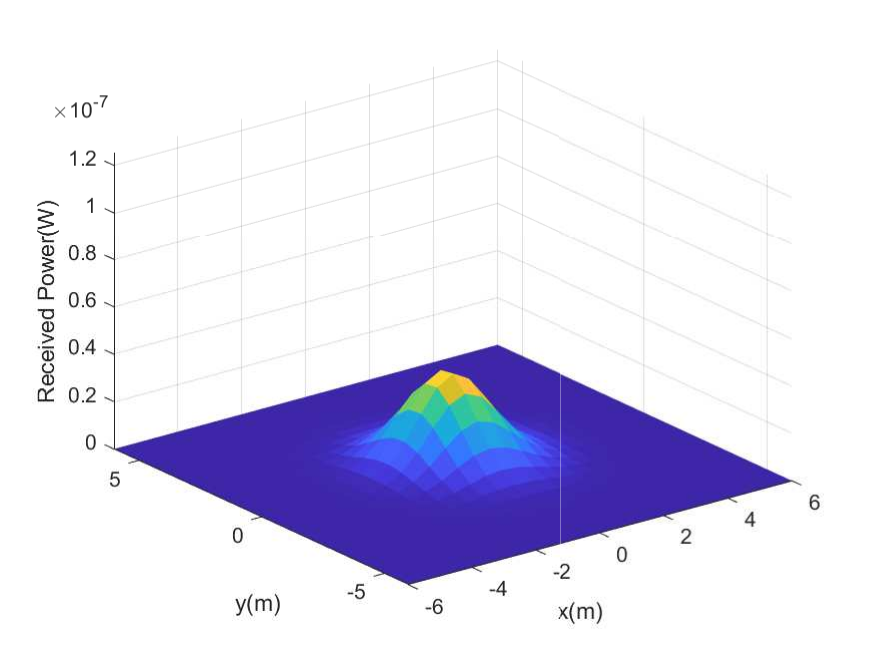}
\label{p_wind_angle}}
\subfigure[]{\includegraphics[width=0.3\textwidth]{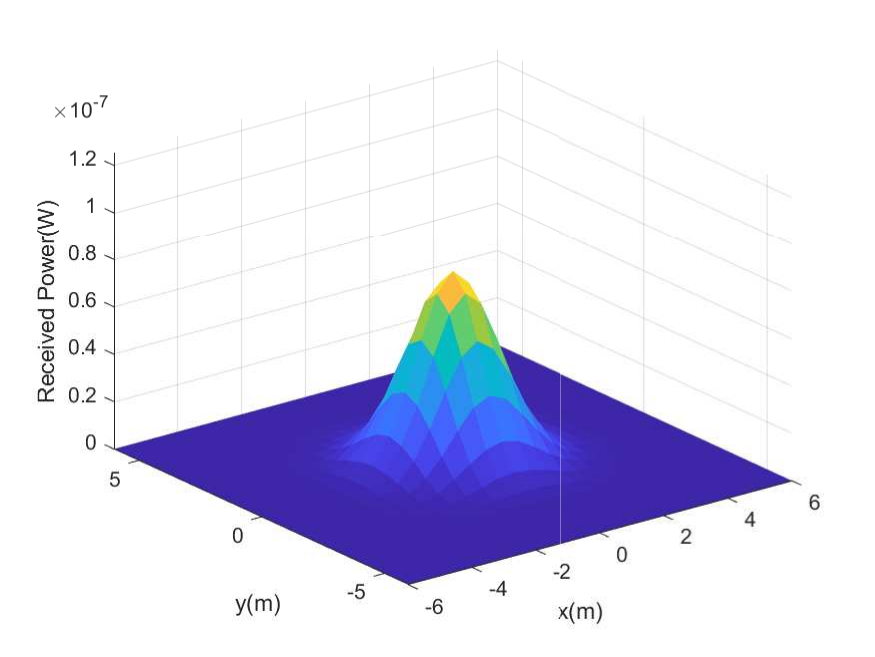}
\label{p_at}}
\caption{Average received power of StULC links: (a) sea surface wind speed $v=6$ $\mathrm{m/s}$, transmit angle $\zeta=0^\circ$, weak atmospheric turbulence; (b) sea surface wind speed $v=12$ $\mathrm{m/s}$, transmit angle $\zeta=0^\circ$, weak atmospheric turbulence; (c) sea surface wind speed $v=6$ $\mathrm{m/s}$, transmit angle $\zeta=30^\circ$, weak atmospheric turbulence; (d) sea surface wind speed $v=12$ $\mathrm{m/s}$, transmit angle $\zeta=30^\circ$, weak atmospheric turbulence; (e) sea surface wind speed $v=6$ $\mathrm{m/s}$, transmit angle $\zeta=0^\circ$, strong atmospheric turbulence}
\label{pre}
\end{center}
\end{figure*}

In this section, we present numerical results of communication performance for StULC system under different environmental conditions. Unless otherwise specified, we set height of satellite $H=200$ km, depth of receiving plane $D=10$ m, transmit zenith angle $\zeta=0^\circ$, wavelength of laser $\lambda=532$ nm, beam divergence angle $\beta_{T}=22$ $\mathrm{\mu rad}$, transmit power $P_{\mathrm{Tx}}=5$ W, atmospheric transmittance $\xi_t=0.7$, the side length of receiving grid cell above the sea surface $d_m=0.1$ m, receiving aperture area $A_r=1.77$ $\mathrm{cm}^2$, receiving zenith angle $\theta_R=90^\circ$, receiving azimuth angle $\phi_R=90^\circ$, receiving FOV angle $\beta_R=90^\circ$, detector responsivity $R_d=0.7$, wind speed at air-water interface $v=6$ $\mathrm{m/s}$; $\epsilon$ is the rate of dissipation of kinetic energy per unit mass of fluid; $\chi_T$ is the rate of dissipation of mean-squared temperature; and $\omega$ is the ratio of temperature and salinity contributions to the refractive index spectrum. According to \cite{elamassie2020vertical}, in regions with depth variations smaller than $30$ $\mathrm{m}$, temperature and salinity changes of seawater are negligible. Therefore, we assume the underwater optical turbulence strength remains constant with depth. For weak underwater turbulence, we set $\epsilon=10^{-2}\mathrm{m}^{2}\mathrm{s}^{-3}$, $\chi_T=10^{-5}\mathrm{K}^{2}\mathrm{s}^{-1}$, and $\omega=-3$; while for strong underwater turbulence, we set $\epsilon=10^{-3}\mathrm{m}^{2}\mathrm{s}^{-3}$, $\chi_T=10^{-4}\mathrm{K}^{2}\mathrm{s}^{-1}$, and $\omega=-0.25$, where $\epsilon$ is the rate of dissipation of kinetic energy per unit mass of fluid. Besides, for clear ocean water, we set $k_a=0.069\mathrm{m}^{-1}$, $k_s=0.080\mathrm{m}^{-1}$, and $g=0.8708$; and for coastal ocean water, we set $k_a=0.088\mathrm{m}^{-1}$, $k_s=0.216\mathrm{m}^{-1}$, and $g=0.9470$.

We first present the spatial distribution of average received power at the receiving plane under varying environmental conditions, considering the wind speed at the air-water interface, atmospheric turbulence, and laser transmit angle in Fig.~\ref{pre}. The spatial power distribution exhibits an approximately radially symmetric Gaussian profile with maximum intensity at the center.
Notably, comparative analysis of Fig.~\ref{p_compare} and \ref{p_wind} at different wind speed reveals a marginal reduction in received power with increasing wind speed. It is because although sea waves induce angle deflections in laser refraction, the deviated beam directions are still concentrated around $\boldsymbol{T'}$. Even with an increase wind speed at the air-water interface, this spatial distribution characteristic will persist. This inherent directioanality indicates that the laser beam has not diverged severely from the random rough sea surface, resulting only a small attenuation of received power.
Fig. \ref{p_angle} and Fig. \ref{p_wind_angle} under different wind speed with trans angle $\zeta=30^\circ$ further comfirm that increasing wind speed at the air-water interface demonstrates negligible impact on received power attenuation, even with varying incidence angles.
Figs. \ref{p_compare} and \ref{p_at} under different atmospheric turbulence strength demonstrate decreased central power and expanded spatial extent of received power, resulting in a flattened distribution profile. This phenomenon stems from atmospheric turbulence-induced beam wander and spreading effects that increase beam radius, consequently redistributing laser energy from the central region toward peripheral areas according to laser principles.
Comparing Fig.~\ref{p_compare} and \ref{p_angle} under different transmit zenith angle, we can observe that as the transmit angle increases, the received power decreases while the receiving range expands. This occurs because a larger transmit angle alters the relative position between the satellite and the receiving plane, making it more difficult for the laser beam to directly reach the receiver. Additionally, the increased transmission distance leads to greater beam divergence, resulting in an expansion of the receiving range.

\begin{figure*}
\centering
\subfigure[]{\includegraphics[width=0.45\textwidth]{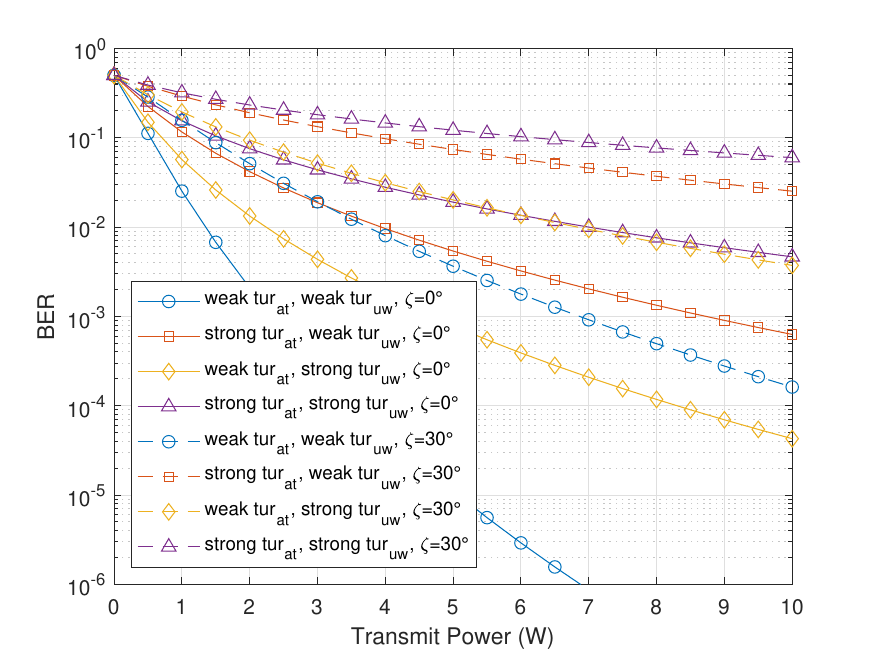}
\label{ber_wind0water0}}
\subfigure[]{\includegraphics[width=0.45\textwidth]{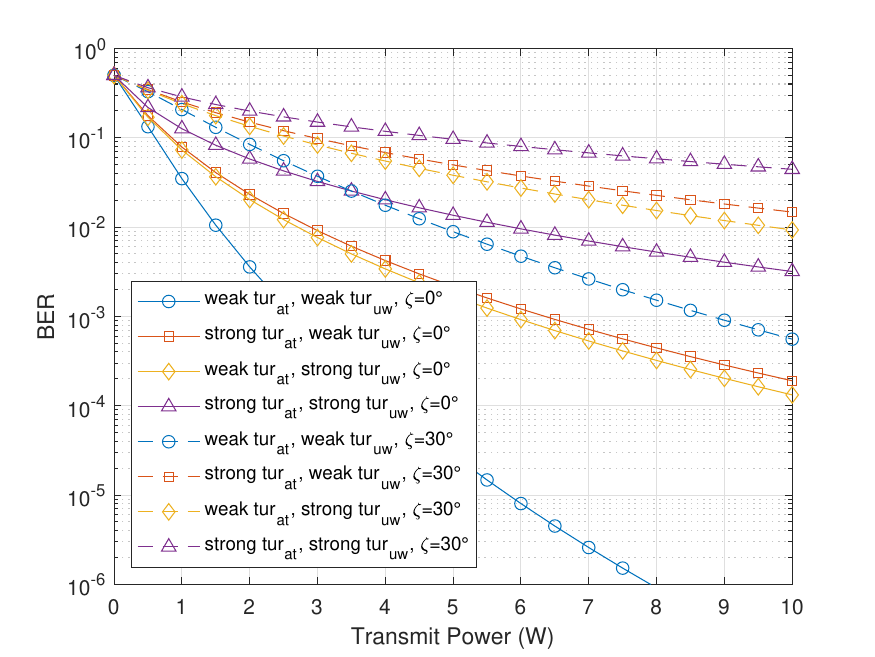}
\label{ber_wind1water0}}
\subfigure[]{\includegraphics[width=0.45\textwidth]{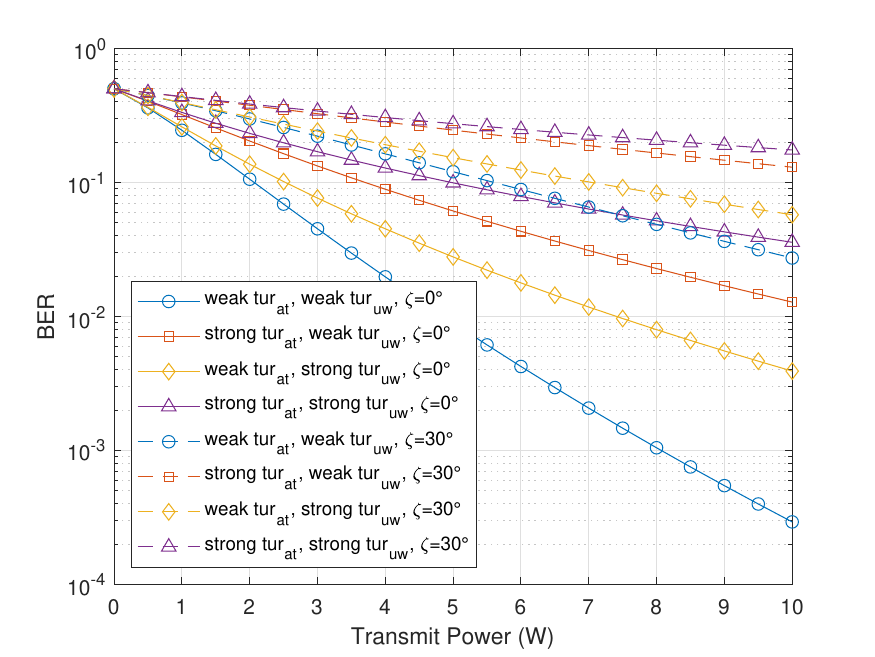}
\label{ber_wind0water1}}
\subfigure[]{\includegraphics[width=0.45\textwidth]{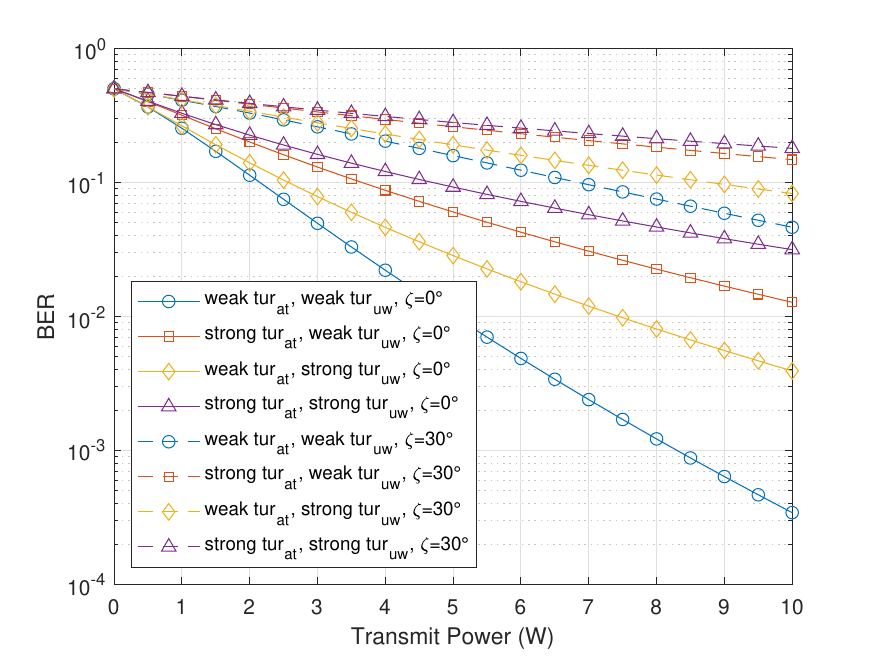}
\label{ber_wind1water1}}
\caption{BER performance of StULC links: (a) clear ocean with $v=6$ $\mathrm{m/s}$; (b) clear ocean with $v=12$ $\mathrm{m/s}$; (c) coastal ocean with $v=6$ $\mathrm{m/s}$; (d) coastal ocean with $v=12$ $\mathrm{m/s}$ }
\label{ber}
\end{figure*}

\begin{figure*}
\centering
\subfigure[]{\includegraphics[width=0.45\textwidth]{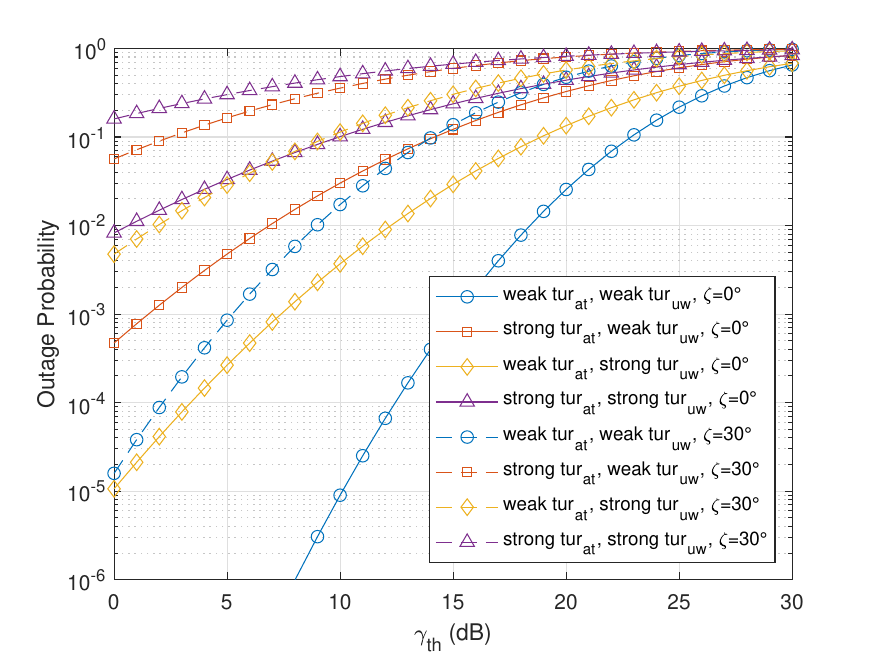}
\label{out_wind0water0}}
\subfigure[]{\includegraphics[width=0.45\textwidth]{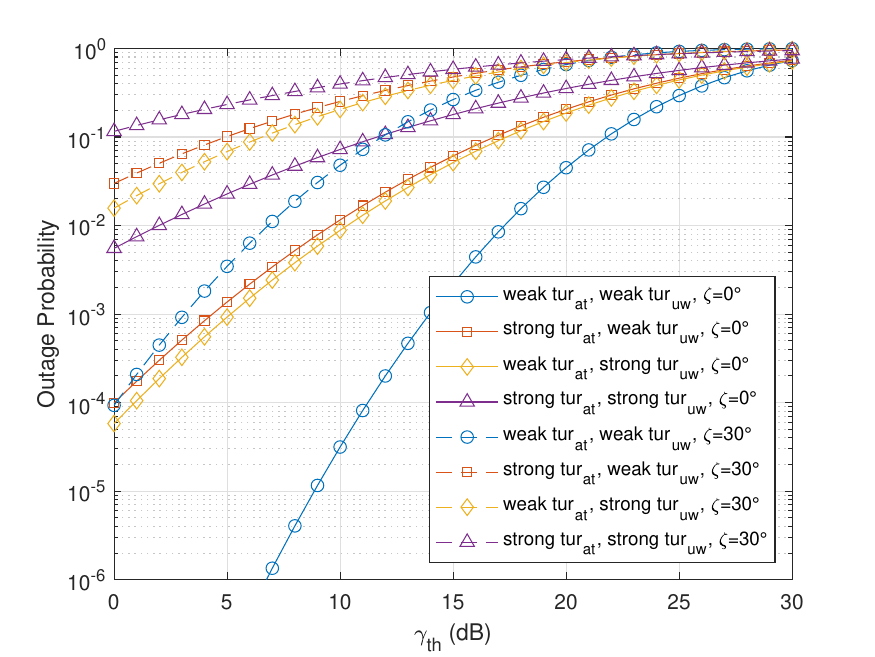}
\label{out_wind1water0}}
\subfigure[]{\includegraphics[width=0.45\textwidth]{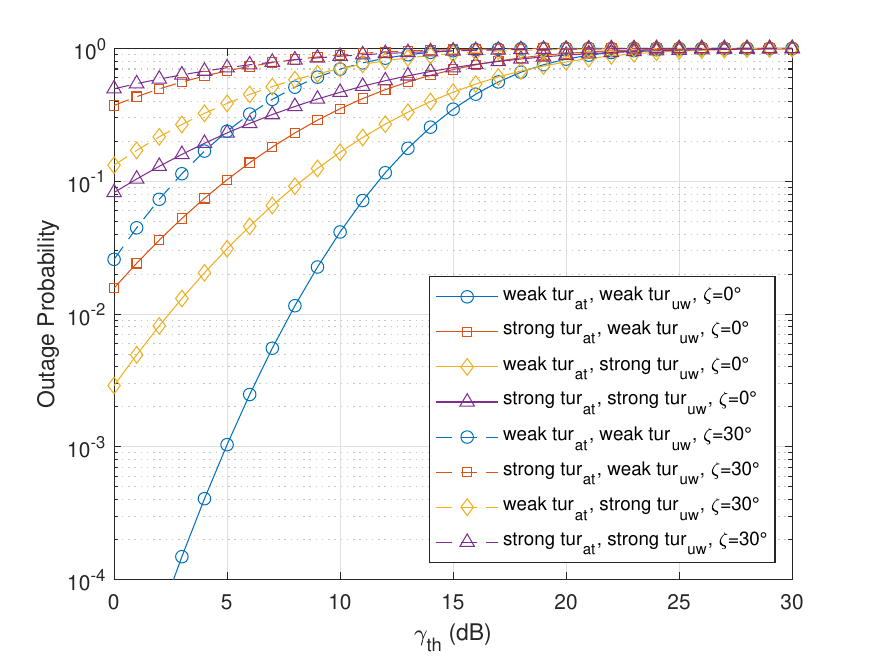}
\label{out_wind0water1}}
\subfigure[]{\includegraphics[width=0.45\textwidth]{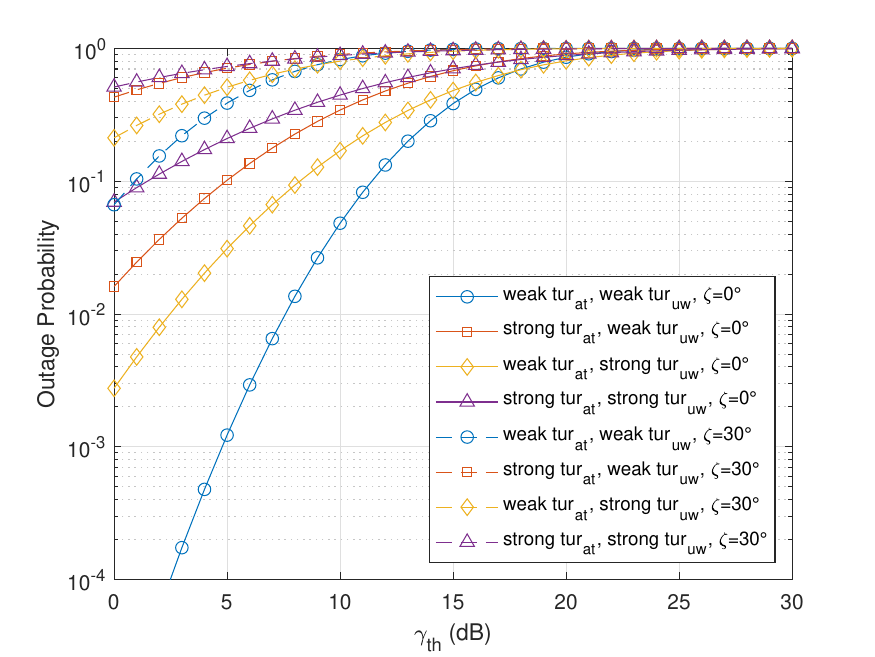}
\label{out_wind1water1}}
\caption{Outage performance of StULC links: (a) clear ocean with $v=6\mathrm{m/s}$; (b) clear ocean with $v=12\mathrm{m/s}$; (c) coastal ocean with $v=6\mathrm{m/s}$; (d) coastal ocean with $v=12\mathrm{m/s}$ }
\label{out}
\end{figure*}

Then we present the BER for different atmospheric and underwater turbulence on received power at varying wind speeds at the air-water interface, underwater particle concentrations, and incidence angles in Fig.~\ref{ber}. From Fig.~\ref{ber} we can find that under identical conditions, BER performance gets worse with higher turbulence strength. Notably, atmospheric turbulence exhibits greater impact on BER than underwater turbulence. This occurs because laser beams propagates significantly farther through the atmospheric channel than the underwater channel, making it more severely affected by atmospheric turbulence.
We can also find that communication performance generally becomes worse with increasing transmit angles. This deterioration primarily stems from extended propagation distance which exacerbates the impact of atmospheric turbulence and reduces the received power.
Comparing Fig.~\ref{ber_wind0water0} and \ref{ber_wind1water0} under different wind speeds, we can observe that BER performance gets worse slightly with increasing sea surface wind speed. This is because the wind speed has limited influence on the receiving power, as we have demonstrated in Fig.~\ref{p_compare} and \ref{p_wind}. Comparative analysis of Fig.~\ref{ber_wind0water0} and \ref{ber_wind0water1} under different water concentrations, we can find that for the coastal seawater conditions with high particle concentration, BER performance exhibits marked deterioration, demonstrating higher sensitivity to particle concentration variations than to turbulence intensity. This is because enhanced scattering effects in coastal ocean will significantly reduce the probability of photons reaching the receiver and thereby attenuate received power.
As further demonstrated again Fig.~\ref{ber_wind0water1} and \ref{ber_wind1water1}, BER performance in coastal ocean exhibits minimal sensitivity to wind speed variations.

At last, we present the outage probabilities for different atmospheric and underwater turbulence on received power at varying sea surface wind speeds and underwater particle concentrations in Fig.~\ref{out}. It is readily seen from Fig.~\ref{out_wind0water0} that higher turbulence strength and larger transmit angle correspond to higher outage probabilities. In addition, comparative analysis of Fig.~\ref{out_wind0water0} and \ref{out_wind1water0} reveals that outage probabilities increase marginally with rising sea surface wind speed. Furthermore, Fig.~\ref{out_wind0water0} and Fig.~\ref{out_wind0water1} demonstrate a significantly increased outage probability in coastal ocean.
These outcomes align with the conclusions drawn in our BER performance analysis.

\section{Conclusion}\label{conclusion}
We established a comprehensive StULC channel model that considered various effects from the atmospheric channel, the air-water interface channel, and the underwater channel.
An analytical method was employed to model the atmospheric channel, avoiding the low computational efficiency caused by long-distance transmission. Besides, a closed-form PDF of the zenith angle of laser propagating direction induced by refractions of random rough sea surface was derived based on the Cox-Munk model, which greatly improves computational efficiency compared to traditional Monte Carlo methods for modeling the air-water interface channel.
The Monte Carlo method was used to model the underwater channel while considering the effects of both underwater particles and turbulence on laser propagation. By utilizing an analytical-Monte Carlo hybrid approach, our model has achieved a balance between computational efficiency and accuracy.
Based on the proposed StULC channel model, spatial distribution of received power, BER, and outage probability were analyzed under different environmental conditions. Numerical results revealed that the decrease in communication performance caused by underwater particle concentration greatly exceeds that caused by atmospheric or underwater turbulence. Besides, numerical results demonstrated that the influence of wind speed at the air-water interface on the communication performance of the StULC link is limited.
Our work shed a light on the research and envelopment of StULC systems.
\bibliographystyle{IEEEtran}
\bibliography{bibdata}

\end{document}